\documentclass[12pt]{iopart}
\usepackage[dvips]{color}
\usepackage[dvips]{epsfig}
\usepackage{cite}
\usepackage{mathptmx}

\begin{document}

\title{Laboratory soft x-ray emission due to the Hawking--Unruh effect?}

\author{G Brodin$^{1,2}$, M Marklund$^{1,2}$, R Bingham$^3$, J Collier$^4$, R G Evans$^{4,5}$}

\address{1 Department of Physics, Ume\aa\ University, SE--901 87 Ume\aa,
Sweden}
\address{2 Centre for Fundamental Physics, Rutherford Appleton Laboratory, 
  Chilton, Didcot, Oxon OX11 OQX, U.K.}
\address{3 Space Science \& Technology Department, Rutherford Appleton Laboratory, 
  Chilton, Didcot, Oxon OX11 OQX, U.K.}
\address{4 Central Laser Facility, Rutherford Appleton Laboratory,
  Chilton, Didcot, Oxon OX11 OQX, U.K.}
\address{5 Physics Department, Imperial College, London, U.K}

\date{Sumitted October 8, 2007}

\begin{abstract}
The structure of spacetime, quantum field theory, and thermodynamics are all
connected through the concepts of the Hawking and Unruh temperatures. The
possible detection of the related radiation constitutes a fundamental test
of such subtle connections. Here a scheme is presented for the detection of
Unruh radiation based on currently available laser systems. By separating
the classical radiation from the Unruh-response in frequency space, it is
found that the detection of Unruh radiation is possible in terms of soft
x-ray photons using current laser-electron beam technology. The experimental
constraints are discussed and a proposal for an experimental design is
given.
\end{abstract}

\maketitle

\section{Introduction}
The discovery that there is a deep connection between black hole physics and
seemingly quite different topics, from information theory and statistical
mechanics to the search for a quantum theory of gravity, has been highly
nontrivial and unexpected. Indeed, it touches upon some of the most
important questions in physics today. In this respect, the Hawking radiation 
\cite{Hawking74,Hawking75} is one of the most interesting processes
predicted in physics, linking general relativity, quantum field theory and
thermodynamics in one single phenomenon \cite{Susskind06}. In contrast to
classical general relativity, a black hole is predicted to loose energy by
thermal radiation with a Hawking temperature 
\begin{equation}
T_{H}=\frac{\hbar g}{2\pi k_{B}c}  \label{Hawking-temp}
\end{equation}
where $k_{B}$ is Boltzmann's constant, $g$ is the horizon surface gravity, $%
\hbar $ is Planck's constant, and $c$ is the speed of light in vacuum. This
result tied in naturally to the concept of black hole entropy presented
earlier by Bekenstein \cite{Bekenstein73}, laying the foundation for black
hole thermodynamics. Since the early discoveries of Bekenstein and Hawking
numerous papers on black hole thermodynamics have been written, see e.g.
Ref. \cite{Wald} and references therein. Closely related to Hawking
radiation is the Unruh-effect \cite{Davies75,Unruh76,Davies-Fulling} (also
known as the Unruh-Davies-Fulling effect), which predicts that zero-point
vacuum fluctuations can be measured as a heat bath of \textit{real particles%
} by an accelerated particle detector. The temperature of this heat bath is
given by the Unruh temperature 
\begin{equation}
T_{U}=\frac{\hbar a}{2\pi k_{B}c}  \label{Unruh-Davies-temp}
\end{equation}
where $a$ is the proper acceleration of the particle detector.
Heuristically, (1) and (2) are related via the equivalence principle, thus
the name Hawking--Unruh effect.

For the large black hole masses expected in astrophysics, the Hawking
temperature (1) is extremely small, and there is evidently no possibility of
ever measuring it. By contrast, several suggestions to measure the
Unruh-effect have been made. For example, Bell and Leinaas \cite
{Bell-Leinaas} suggested to look for the Unruh effect in storage rings,
Ref.\ \cite{Rogers} presented the Penning trap as a possible means for
detecting the Unruh temperature, Ref.\ \cite{Yablonovitch} discussed the
possibility of using a rapidly changing refractive index for detection of
Unruh radiation, and Ref.\ \cite{Schutzhold-Unruh} proposed, based on the
concept of analogue gravity, the use of electromagnetic waveguides to
simulate the effects of curved spacetime quantum field theory. Although
there are numerous works confirming the concept of the Unruh effect \cite
{Davies75,Unruh76,Obadia-Parentani-I,Obadia-Parentani-II,Davies2005} , there
is still much controversy surrounding this issue \cite
{Ford-Conell,Grove86,Narozhny2004}, thus making it even more important to
gain experimental or observational insight into these concepts. For example,
Narozhny \textit{et al}. argues that the heat bath is not a universal
consequence of acceleration \cite{Narozhny2004}. On the other hand, Ford and
O'Conell \cite{Ford-Conell} presents results for a scalar field which
supports the concept of a heat bath eventually leading to thermalization at
the Unruh temperature for an accelerated system. However, Ford and O'Conell
conclude that the system will not emit radiation that can be detected in
the laboratory frame, in spite of its finite temperature. For a different
opinion concerning the reality of the Hawking--Unruh radiation and further
discussions of this issue, see also Refs. \cite{Unruh92,Fulling-Unruh2004}.
Assuming that the Unruh-radiation is real, it is clear that a successful
detection would be an important milestone. Unfortunately the Unruh-signal is
drowned in noise in most of the early proposals. A setup which tries to
address this problem has been suggested by Ref. \cite{Chen-Tajima}, where
strong lasers are used to accelerate electrons to ultra-relativistic
energies, corresponding to Unruh-temperatures $T_{U}=10^{4}\,\mathrm{K}$, or
higher. While the heat bath felt by the accelerated electrons only
contributes with a small correction to the electron motion, the ordinary
Larmor radiation has a ''blind spot'' in the direction of acceleration. Thus
the idea was that properly placed detectors may measure radiation produced
by electrons responding to the heat bath, without being drowned in the
standard Larmor radiation. Similar ideas were put forward in Ref. \cite
{Schutzhold2006}, where modifications due to photon pair correlation and
field strengths approaching the Schwinger critical field also were included. 

In the present work we propose an experimentally viable method for the
detection of the Hawking--Unruh effect using currently available technology.
We consider electrons accelerated by lasers in a novel geometric set-up as a
means for producing a detectable signal due to the accelerated vacuum
temperature. In particular, we chose a configuration with circular electron
orbits. This leads to an energy distribution of the vacuum fluctations, in
the electron rest frame, that to some extent deviate from a thermal heat
bath. However, the results of for example Ref. \cite{Korsbakken-2004} for
circular motion, shows that an effective temperature of the vacuum
fluctuations can still be defined, such that Eq. (\ref{Unruh-Davies-temp})
continues to be a useful approximation. This allows us to calculate the
momentum distribution function for the generated non-classical photons.
Since a relatively high electron number density is needed for appreciable
emission, we include collective effects when comparing with the classical
emission. It turns out that the classical power is orders of magnitude
larger than the Unruh power, for experimental parameters currently
available, a result which is highly problematic in other experimental
suggestions. However, here the calculated distribution function shows a
clear signature distinguishing Unruh photons from classical photons, in
particular giving rise to photons in the soft x-ray regime. We analyse the
effects from competing sources, and we conclude that using laser-electron
beam systems \textit{currently} in operation it is possible to obtain a
clearly detectable signal.

\section{Theory} 

\subsection{The Unruh effect in laser fields}
We will rely on ultra-intense lasers to accelerate electrons. However, due
to the unfavorable scaling with the electron number density $n_{e} $, of the
amount of radiation produced, we deduce that in a setup with a reasonably
high electron density, the classical power will always be much larger than
the Unruh-radiation. In order to separate the Unruh-response from the
classical radiation we therefore compare their spectral profiles. For this
purpose we first note that for presently available field strengths $E_{0}$,
we have $E_{0}\ll E_{\mathrm{crit}}$, where $E_{\mathrm{crit}} =
m^2c^3/e\hbar \approx 10^{16} \, \mathrm{V/cm}$ is the Schwinger critical
field. Here $m$ is the electron rest mass and $e$ is the elementary charge. From
Eq. (\ref{Unruh-Davies-temp}) we thus see that the photon energies of the
heat bath will be much smaller than the electron rest mass. Accordingly, in
order to describe the interaction of the electrons with the heat bath, we
will view the process as Thomson scattering in the rest frame of the
electrons. In essence this means that the detection of the Unruh effect is
made by accelerated electrons acting as spherical mirrors of virtual
photons, making them real in the laboratory frame. This picture of the
process is given further support by the results of Refs. \cite
{Obadia-Parentani-I,Obadia-Parentani-II}, where the analog between moving
mirror radiation and Unruh radiation is put on a firm ground.

Introducing the distribution function of thermal photons $f_{B}(\mathbf{k}%
,T_{U})$, 
the rate of scattered photons $d\mathcal{N}_{s}/d\tau $ by a single electron in the
rest frame can be written 
\begin{equation}
\frac{d\mathcal{N}_{s}}{d\tau }=\frac{d}{d\tau }\int f_{s}k^{2}d\Omega dk=\sigma
_{T}c\int f_{B}k^{2}\,d\Omega \,dk  \label{Number-rate}
\end{equation}
where $k$ is the photon wavenumber, $\sigma _{T}=e^{4}/(6\pi \varepsilon
_{0}^{2}m^{2}c^{4})$ is the total Thomson scattering cross section, $f_{s}$
is the distribution function of the scattered photons, and $\varepsilon _{0}$
is the vacuum permittivity and $\tau $ is the proper time in the electron
rest frame. 
Next we represent the distribution function as a locally thermal radiation
distribution $f_{B}(\mathbf{k,}T\mathbf{(r))}$ and write the number of
scattered photons $N_{s}$ from a volume $V$ such as 
\begin{equation}
\frac{dN_{s}}{d\tau }=\int_{V}n_{e}(\mathbf{r})\sigma _{T}c\int f_{B}(%
\mathbf{k,}T\mathbf{(r))}k^{2}\,d\Omega \,dk\,dV  \label{number-rate-II}
\end{equation}
where $n_{e}$ is the number density of electrons. Hence the scattered
distribution function evolves according to $df_{s}/d\tau =\sigma _{T}cf_{B}$. 
Next we assume that the volume is sufficiently small, such that for a
given time the spatial dependence of the electron velocity is negligible
within $V$ \ (i.e. we only count the contribution from the central part of
the laser pulse ). The power from the Unruh effect $P_{U,\mathrm{rest}}$
emitted in the electron rest frame then reads 
\begin{equation}
P_{U,\mathrm{rest}}=\frac{d}{d\tau }\int_{V}\int \hbar \omega _{\mathrm{rest}%
}f_{s}(\mathbf{k})k^{2}\,d\Omega \,dk\,dV  \label{Power-rest}
\end{equation}
where $\omega _{\mathrm{rest}}$ is the photon frequency in the rest frame.
In order to evaluate the power $P_{U,\mathrm{lab}}$ emitted in the
laboratory frame, we introduce spherical coordinates, with the $z$-axis
perpendicular to the velocity, and we write 
$\omega _{\mathrm{lab}}=\omega _{\mathrm{rest}}\gamma (v)\left[ 1-(v/c)\sin
\theta \cos (\phi -\phi _{v})\right] $, 
where $\omega _{\mathrm{lab}}$ is the photon frequency in the lab-frame and $%
\phi _{v}$ is the angle between the velocity and the $x$-axis. Using the
analogy between moving mirror radiation and the Unruh effect \cite
{Obadia-Parentani-I,Obadia-Parentani-II}, the emitted laboratory power is
then calculated as 
\begin{equation}
P_{U,\mathrm{lab}}=\int_{V}\int \hbar \omega _{\mathrm{lab}}\frac{df_{s}}{%
d\tau }\frac{d\tau }{dt}k^{2}\,d\Omega \,dk\,dV=P_{U,\mathrm{rest}}
\label{Power-lab}
\end{equation}
where the last step comes from noting that 
$f_{s}$ and the phase space volume element are scalars.

Next, we want to compare the radiation generated due to the Unruh effect
with the classical emission. As a model, we consider electrons accelerated
by counter propagating laser pulses with left and right hand circular
polarization. The advantage with circular polarization is twofold. Firstly,
due to the high symmetry, a simple harmonic current response with circular
orbits solves the fluid equations of motion \cite{Stenflo1976}, and thus the
electron response is more easily evaluated. Secondly, collective nonlinear
effects (e.g. harmonic generation, induced density fluctuations, etc., see
Ref. \cite{Shukla1986} for a list of possible mechanisms) that may induce
classical competing high-frequency emission is thereby minimized. The
electric field can then be written $
\mathbf{E}=E_{0}\left[ \exp i(k_{0}z-\omega _{0}t)+i(-k_{0}z-\omega _{0}t)%
\right] \widehat{\mathbf{x}}/2+E_{0}\left[ \exp i(k_{0}z-\omega
_{0}t)+i(-k_{0}z-\omega _{0}t)\right] \widehat{\mathbf{y}}/2+\mathrm{c.c.},$ 
where $\mathrm{c.c.}$ stands for complex conjugate and $\omega _{0}$ and $%
k_{0}$ are the laser frequency and wavenumber, respectively. We note that in
a region $\left| z\right| \ll \lambda _{L}$, where $\lambda _{L}$ is the
laser wavelength, the corresponding magnetic field is vanishingly small for
all times. Concentrating on this region, the electron velocities can be
written as 
\begin{equation}
\mathbf{v}=\frac{eE_{0}\left[ \cos (\omega _{0}t)\widehat{\mathbf{x}}+\sin
(\omega _{0}t)\widehat{\mathbf{y}}\right]}{m\omega _{0}\sqrt{%
1+(eE_{0}/m\omega _{0}c)^{2}}}  \label{velocity}
\end{equation}

The Unruh emission is described by the scattered distribution function $%
f_{s}(\mathbf{k})$, which is shown in Fig.\ 1 for $\gamma =70$. We note the
strong beaming in the direction of the velocity. 
We deduce from Eqs.\ (\ref{Number-rate})--(\ref{Power-rest}) and (\ref
{velocity}) that the energy distribution due to the Unruh effect in the
laboratory frame, averaged over a period time, will be (approximately)
thermal with a temperature 
\begin{equation}
T_{\mathrm{lab}}=\frac{\gamma e\hbar E_{0}}{2\pi kcm}  \label{Lab-temp}
\end{equation}
in spite of the strong anisotropy of $f_{s}$. The high temperature means
that the characteristic wavelength in the laboratory frame typically will be
much shorter than the laboratory orbit radius, a result which will be
helpful in separating the Unruh radiation from the classical radiation. Here
it should be noted that in the case that the particle deviates from
stationary orbits, the asymptotic part of the spectra will follow a power
law, rather than being thermal, as deduced in Ref. \cite{Obadia-Milgrom}.
Since, such deviations are inevitable, the highest frequencies will
certainly not be thermally distributed. Our detection scheme, however, is
based on detection of the central part of the frequencies. Thus a small
deviation from stationarity is of limited importance in our case.
Here we also neglect the effect that the photons in the heat bath are
created in pairs. Note, however, that a proposal that attempts to use this
fact, measuring the correlation between individual scattering events, has
recently been proposed \cite{Schutzhold2006}.

Next evaluating the radiated Unruh power 
using Eqs.\ (\ref{Number-rate})--(\ref{Power-rest}) we obtain 
\begin{equation}
P_{U,\mathrm{lab}}=\frac{Ne^{8}E_{0}^{4}\hbar }{1440\pi
^{3}c^{10}m^{6}\varepsilon _{0}^{2}}  \label{Power-Unruh}
\end{equation}
We note that even for the highest laser fields currently available, $%
E_{0}\sim 2\times 10^{12}\,\mathrm{V/cm}$, we still need a large number of
accelerated electrons (at least $N>10^{8}$) to reach detectable levels of
the Unruh radiation.

\subsection{Spectral structure}
Let us now study the spectral properties of the radiation. For an irradiance
of $10^{21} - 10^{22}\,\mathrm{W/cm^{2}}$, we note that the characteristic energy of
the Unruh photons [cf.\ (\ref{Lab-temp})] is larger than the laser
photon energies, by several orders of magnitude,
assuming the laser works in the optical range. Thus the assumption to view
the periodic variation of the laser field as slow when considering the
spectral properties of the Unruh radiation is justified. Furthermore, within
the framework of our model, the spectral properties of the classical emission is
suppressed for a wide range of frequencies (see Sec.\ 3), and
thus it is straightforward to separate the two contributions. However, we note that
we will only get a few photons due to the Hawking--Unruh effect per laser
shot, even for a rather high electron number density, $n_{e}=10^{21}\,%
\mathrm{cm}^{-3}$. Combining Eqs.\ (\ref{Lab-temp})--(\ref{Power-Unruh}) we
can write the number of photons per shot as 
\begin{equation}
  N_{U}=0.084\left(\frac{E_{p}}{15\,\mathrm{J}}\right)^{2}\left(\frac{0.5\times 10^{15}\,\mathrm{W}}{P}\right)\left(\frac{800\,\mathrm{nm}}{\lambda_{L}}\right)\left(\frac{n_{e}}{10^{21}\,\mathrm{cm}^{-3}}\right), 
\end{equation}
where $E_{P}$ is the pulse energy
for each pulse, $P$ is the laser power, $\lambda _{L}$ is the laser
wavelength and $n_{e}$ is the electron density. Similarly, the
characteristic photon energy can be written as 
\begin{equation}
  \hbar \omega _{\mathrm{char}}=582\left(\frac{I}{10^{22}\,\mathrm{W\,cm^{-2}}}\right)\left(\frac{\lambda _{L}}{800\,\mathrm{nm}}\right) ,
\end{equation}
where 
$I$ is the intensity and $\hbar \omega _{\mathrm{char}}$ is given in units
of eV. In Fig.\ 2 a schematic view of the experimental set-up is given. Two
opposed focused laser beams with opposite circular polarization are allowed
to interact with an under-dense laser produced electron beam, with a drift
velocity well below $c$. The soft x-ray photons due to the Hawking--Unruh
effect are emitted in a narrow band perpendicular to the incoming laser
beams. Careful placement of the photon detectors leads to a nearly full
x-ray coverage. 
The output when varying the intensity in a number of different laser systems
(for which the necessary numbers are given in Table 1) is shown in Fig.\ 3.

\section{Classical radiation}
Next we study the vector potential due to the classical emission, that can
be written $\mu _{0}\int (\mathbf{J(}t_{\mathrm{ret}},\left| \mathbf{r-r}%
^{\prime }\right| \mathbf{)}/r)dV^{\prime }$, in the radiation zone, where $%
t_{\mathrm{ret}}$ is the retareded time. A major difficulty for detecting
the Unruh contribution is that the Larmor (or synchrotron) radiation power
(described by the single particle delta function contributions to the
retarded current) is always much larger than the Unruh response for all
frequency regimes. These arguments apply if we consider the radiation from a
single particle. When increasing the particle density, the Unruh radiation
power (\ref{Power-Unruh}) scales linearly with the number of particles, but
the classical radiation may grow equally fast or even faster. Fortunately,
different parts of the classical spectrum behave differently, depending on
whether the emitted radiation is a collective fluid effect, or a single
particle effect. If we look on the shortest emitted wavelengths,
neighbouring particles are typically much more than a wavelength away, and
we cannot detect interference effects in the radiation pattern. As a
consequence, the classical radiation scales linearly with the number of
particles, and it will be extremely difficult to see the Unruh response in
the emitted radiation for such wavelengths, since the gyrating motion of the
electrons produces synchrotron radiation at harmonics of the laser frequency
up to the frequencies of the order $\sim \gamma ^{3}\omega _{0}$. In
principle, for strictly periodic motion the synchrotron radiation is built
up of multiples of the laser frequency, but a finite bandwidth of the laser
will limit our possibility to benefit from this. However, studying radiation
with wavelengths much longer than the nearest neighbour distance of the
particles is more useful. For the electron densities considered, $%
n_{e}=10^{21}\,\mathrm{cm}^{-3}$, wavelengths of the order $\lambda =0.01%
\mathrm{\mu m}$ could be suitable, fulfilling $\lambda \ll L_{nn}$, where $%
L_{nn}$ is the nearest neighbour distance. For such wavelengths, the single
particle radiation is suppressed\footnote{
 The power of singe particle radiation scales
as $1/N$, where $N$ is the number of electrons (this result is obtained from a
random walk model, in which the contribution to the vector potential
from each particle has a random phase compared to the others). However, when
the wavelength of the radiation is much longer than the nearest neighbor
distance, the uncertainty of the phase of the contribution from each
accelerated electron is much less than $\pi /2$, and the single particle
radiation is thereby suppressed.
} 
and the output from
the gyrating motion can be described with a fluid model. Assuming that the
current density decays smoothly with the radius, and using Eq. (\ref
{velocity}) the classical radiated power becomes $P_{\mathrm{classical}%
}=\eta N^{2}e^{4}E_{0}^{2}/m^{2}c^{3}\gamma ^{2}$ 
where $N=n_{e}2\pi R^{2}L$ is the number of electrons in a pulse with length 
$2L$ and radius $R$, and $\eta $ is a geometrical factor of order unity.
This collective fluid response is the main source of classical emission, but
its frequency spectrum is determined by the laser frequency, and thus it is
limited to the optical range. 

The result that the output is negligible for frequencies deviating from the
fundamental frequency depends on two facts. Firstly, the time-dependence 
of the current density in
each volume element should be quasi-monochromatic such that we can write $%
\mu _{0}\int_{V^{\prime }}\mathbf{J}(t_{\mathrm{ret}},\left| \mathbf{r}-%
\mathbf{r}^{\prime }\right| /c)dV^{\prime }=\exp [-i\omega (t-r/c)-]\mu
_{0}\int_{V^{\prime }}\exp (i\omega \mathbf{r\cdot r}^{\prime }/cr^{2})%
\mathbf{J}_{s}(\left| \mathbf{r}-\mathbf{r}^{\prime }\right| )dV^{\prime }$
in the radiation zone, where $\mathbf{J}_{s}(\left| \mathbf{r}-\mathbf{r}%
^{\prime }\right| )$ contains the spatial dependence of the current density.
Secondly, in order to supress synchrotron radiation effects, the volume $%
V^{\prime }$ where the current density $\mathbf{J}_{s}$ is non-zero \textit{%
must not} be strongly time-dependent. This later condition is fulfilled if
the radial dependence of the pulse electric field smootly approaches zero
outside the central pulse region, in which case the weak time-dependence of
the interaction volume also makes the expression $\int_{V^{\prime }}\exp
(i\omega \mathbf{r\cdot r}^{\prime }/cr^{2})\mathbf{J}_{s}(\left| \mathbf{r}-%
\mathbf{r}^{\prime }\right| )dV^{\prime }$ time-independent. As a
consequence the higher frequencies of the classical spectra is suppressed up
to the wavelengths where the fluid model breakes down. Thus, there is a
window of detectable wavelengths, shorter than the laser wavelength, but
much longer than the inter particle distance. The parameters of an
experiment can be chosen to fit the Unruh radiation into this window.

\begin{figure}[tbp]
\centering
\includegraphics[width=0.98\columnwidth]{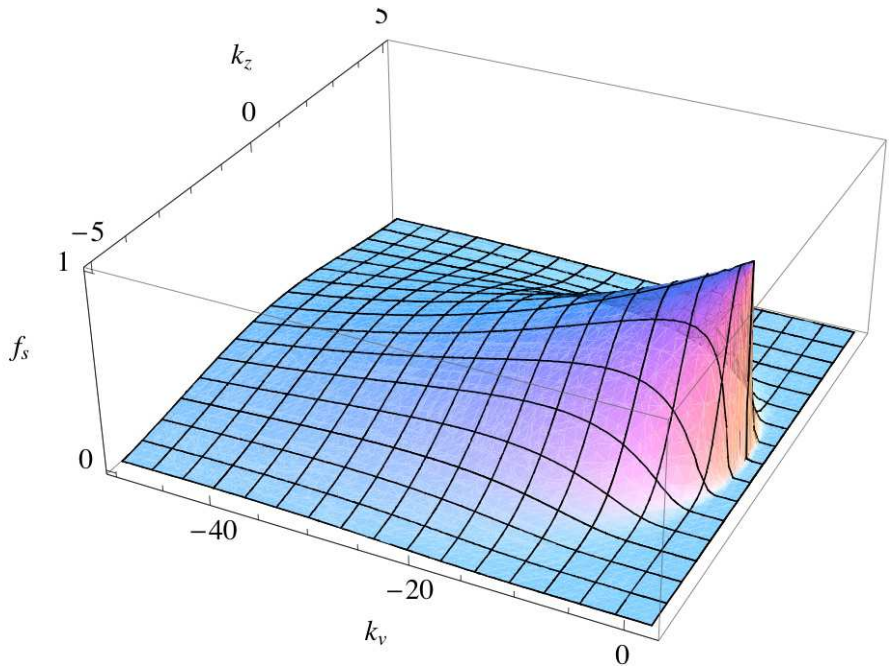}
\caption{A cross section of the normalized scattered distribution function $%
f_{s}(k_{v},k_{z})/f_{s}(0,0)$ as a function of normalized wavenumbers ($%
\hbar k_{v}/k_{B}T_{U}\rightarrow k_{v}$ and $\hbar
k_{z}/k_{B}T_{U}\rightarrow k_{z}$). Here $k_{v}$ denotes the wavenumber in
the direction of the velocity. The figure corresponds to $\protect\gamma %
\approx 70$, and the cross-section is shown for $k_{\bot }=0$, where $%
k_{\bot }$ is the component perpendicular to $k_{v}$ and $k_{z}$. We see
that there is a strong beaming effect in the direction of the velocity.}
\end{figure}

\begin{figure}[tbp]
\centering
\includegraphics[width=.98\columnwidth]{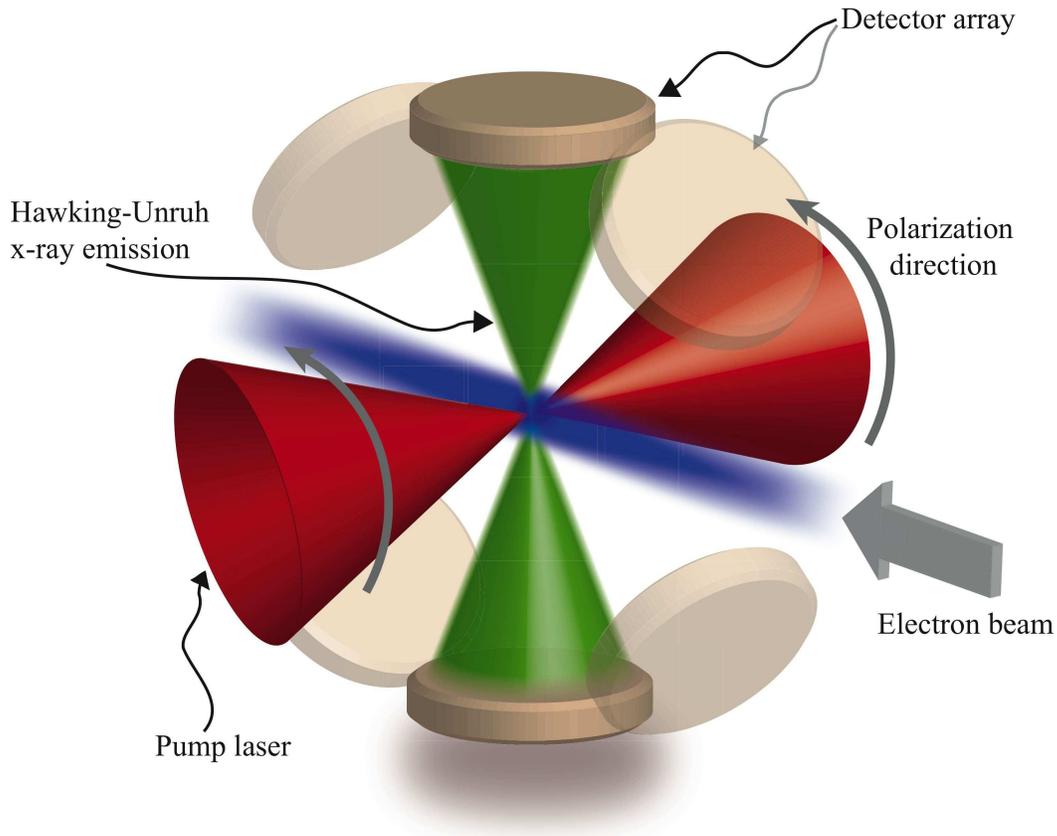}
\caption{A schematic view of the setup. The emitted radiation will be
sharply beamed in the direction of the electron velocity. Thus the detectors
only have to cover a short distance in the direction of propagation}
\end{figure}


\begin{table}[th]
\caption{The relevant parameter values for different two-beam laser systems. In the first column the each 
laser pulse energy is given, in the second the pulse power, in the third the pulse focal intensity, and in the fourth
the laser wavelength. The Ti:Sapphire is assumed to have standard high-intensity properties, the Astra--Gemini system is in operation (from 2007) at the Rutherford Appleton Laboratory (RAL) in the U.K.,
with a proposed upgrade, 
the Vulcan laser is in operation at the RAL, and
the Omega EP laser will be operational in Rochester (USA). The HiPER (High Power 
Experimental Research Facility) and ELI (Extreme Light Infrastructure) are European infrastructure
projects under planning \cite{Dunne,Gerstner}. 
}
\begin{indented}
\item[]
\begin{tabular}{@{}lllll}
\br
  \textbf{Laser type} & \textbf{Energy ($\mathrm{J}$)} & \textbf{Power ($\mathrm{PW}$)} 
    & \textbf{Intensity ($\mathrm{W/cm^2}$)} & \textbf{Wavelength} ($\mathrm{nm}$) \\
\mr
  Ti:Sapphire     & $1$                 & $0.03$ & $10^{21}$              & $800$ \\
  Astra--Gemini & $15$                & $0.5$  & $10^{22}$               & $800$ \\
  AG upgrade    & $15$               & $0.5$   & $\leq 10^{24}$ & $800$ \\
  Vulcan           & $250$              & $0.5$  & $5\times10^{20}$     & $1054$ \\
  Omega EP     & $\geq 2500$ & $0.25$     & $6\times10^{20}$     & $1054$ \\
  HiPER 1        & $4500$            & $150$  & $5\times10^{24}$     & $1054$ \\
  HiPER 2, ELI        & $37500$          & $2500$ & $5\times10^{26}$    & $1054$ \\
\br
\end{tabular}
\end{indented}
\end{table}

\begin{figure}[tbp]
\centering
\includegraphics[width=0.98\columnwidth]{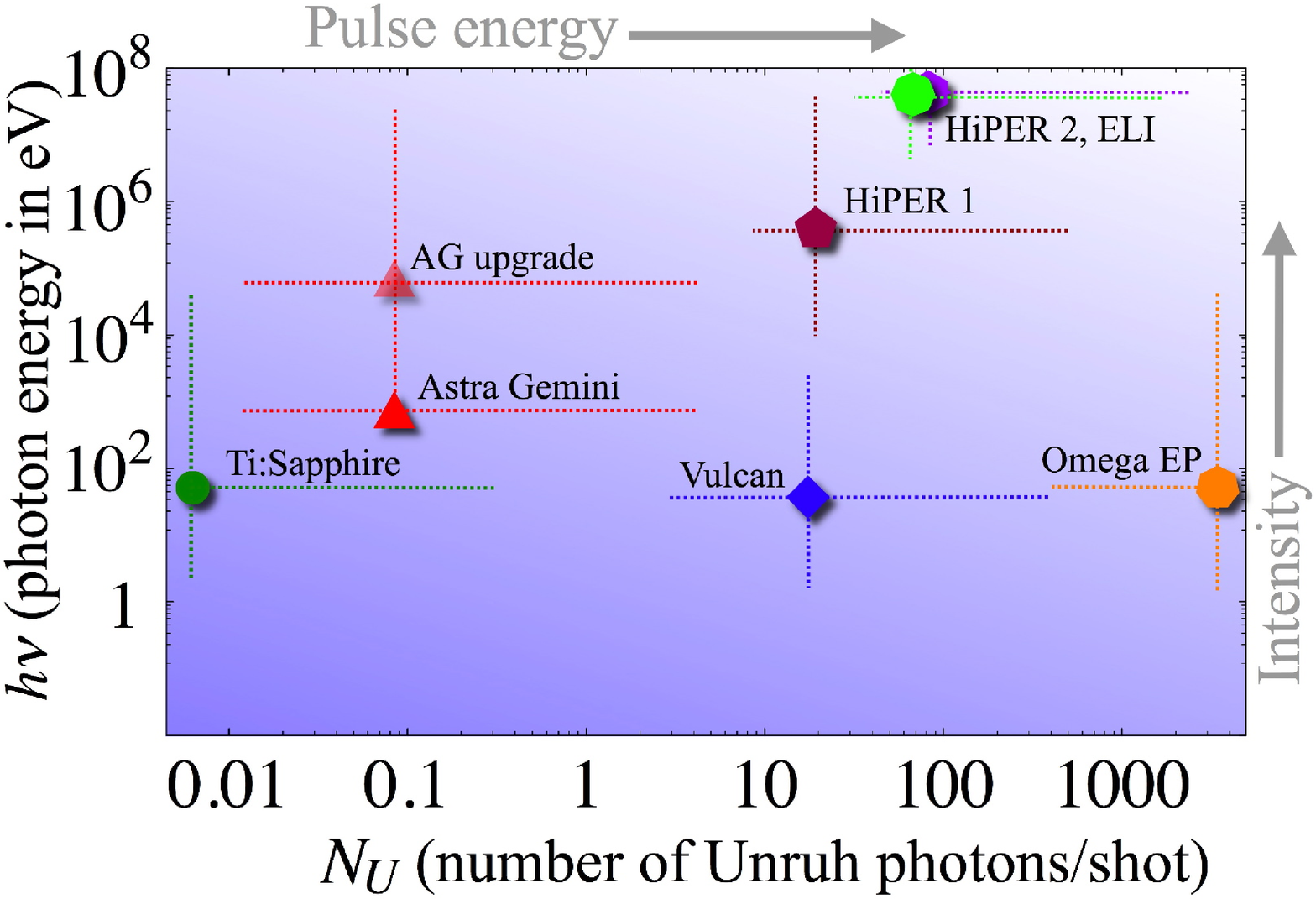}
\caption{ The figure shows the characteristic energy $\hbar\protect\omega_{%
\mathrm{char}}$ in units of eV of the photons generated due to the Unruh
effect and its relation to the number of photons $N_U$ per shot for
different ultra-intense laser systems. The horizontal lines correspond to
variations in pulse energy (pulse length), keeping the focused intensity
constant. The vertical lines correspond to variations in intensity, keeping
the pulse length and pulse energy constant. The different values are
computed using the same electron number density $n_e = 10^{21}\,\mathrm{cm}%
^{-3}$. The relevant parameter values can be found in Table 1. }
\end{figure}
Other mechanisms that could generate radiation competing with the Unruh
contribution includes particle collisions. To avoid this source, we consider
a laser produced electron beam, rather than a plasma, in order to prevent
the high energy electrons to scatter of ions and produce competing soft
x-rays. Moreover, we note that efficient absorption of laser energy with
subsequent x-ray generation \cite{Skobelev2002} due to electron-electron
collisions can be avoided if the density is kept undercritical. This is
consistent with an electron beam density $n_{e}\sim 10^{21}\,\mathrm{cm}^{-3}
$, in which case the synchrotron emission is suppressed for photon energies
up to $1\ \mathrm{KeV}$. 
This leads to an optimal irradiance in the range $\sim 10^{21}-10^{22}%
\mathrm{W/cm^{2}}$. Specifically, for a pulse energy of $3\mathrm{kJ}$, a
focused intensity of $10^{21}\mathrm{W/cm^{2}}$, and a wavelength of $1%
\mathrm{\mu m}$ together with an electron density $10^{21}\,\mathrm{cm}^{-3}$
we generate more than $2\times 10^{3}$ Unruh photons/shot, with an energy of
the order of 100\textrm{eV }. Naturally, the next generation of laser
facitilites in the early planning stage, like ELI or HiPER would produce
even more impressive results. It should be stressed, however, that our
experiment does not benefit from the huge focusing capabilities in those
cases, since very high intensities move the characteristic photon energy
outside the window where classical emission is suppressed. Nevertheless, in
case a proper degree of focusing is chosen, naturally the large pulse
energies in facilities like HiPER 2 or ELI (see table 1)make them excellent
choices for our suggested experiment. 
In conclusion, we find that classical soft x-ray emissions can be
sufficiently suppressed in the relevant parameter regime. 

\section{Conclusion and discussion}
An unambiguous detection of the Hawking--Unruh effect would be an important
breakthrough, shedding light on the deep and fundamental connection
between general relativity, quantum field theory, and thermodynamics and
settling much of the controversies that have arisen over the subject, see
e.g. Refs. \cite{Ford-Conell,Narozhny2004,Unruh92,Fulling-Unruh2004}. The
details of the possible signals detected through such experiments could even
have interesting consequences for a future quantum theory of gravity \cite
{ling2006,amelino-camelia2006}. Noting the similarity with radiation from
moving mirrors \cite{Davies2005,Obadia-Parentani-I,Obadia-Parentani-II}, we
have calculated the distribution function for the photons generated from
laser accelerated electrons, due to the Unruh effect. 
 The reader might ask whether the analogy between infinitely heavy mirrors and finite mass electrons is valid or not. After all, it is the finite mass of the electrons that allow them to respond to the heat bath. However, we stress that when determining the cross-section of the electron "mirrors", the finite electron mass is incorporated. Moreover, Chen \& Tajima \cite{Chen-Tajima} suggested a novel means to detect the Unruh effect through single electron dynamics. However, in practice a large electron density is needed, making collective effects essential. In particular, the competing classical radiation scales as $N^2$, while the Unruh radiation scales as $N$, where $N$ is the number of electrons in the interaction region. Thus, the effective spatial window in the Larmor radiation scales as $1/N$ ($\sim 10^{-9}$ for an electron density of $10^{21}\,\mathrm{cm}^{-3}$ and a spot-size of the order of $\mu\mathrm{m}$), making spatial filtering insufficient. Instead, one needs to include the collective properties of the classical radiation, when comparing the classical and Unruh {\it spectral} distributions.    
For the spectral
signature of the Unruh photons to be distinguishable from competing effects,
the classical soft x-ray emission must be eliminated. The key in doing this
is to consider a pure electron plasma to avoid electron-ion scattering, and
to limit the heating of the laser target, which can be achieved by keeping
the electron density well below the critical density \cite{Skobelev2002}.
For an electron beam density $n_{e}\sim 10^{21}\,\mathrm{cm}^{-3}$ the
classical synchrotron emission is suppressed up to photon energies of $1\,
\mathrm{keV}$ due to destructive interference, which means that the spectral
properties of the Unruh radiation is ideal for detection for an irradiance
of the order $10^{21}-10^{22}\mathrm{W/cm^{2}}$. In conclusion, we have
shown that, through proper experimental design, detection of the Unruh
effect using accelerated electron is possible with currently available
technology.

\section*{Acknowledgements}
  This research was supported by the Swedish Research Council and the Centre
for Fundamental Physics, Rutherford Appleton Laboratory, U.K.

\end{document}